\documentclass[aip,apl,twocolumn,10pt]{revtex4-1}
\usepackage{graphicx}%
\usepackage{dcolumn}%
\usepackage{bm}%
\usepackage{epstopdf}
\usepackage{subfigure}
\usepackage{afterpage}
\usepackage[sort&compress]{natbib}

\draft 

\begin{document}


\title{Polarized Temperature Dependent Raman Study of Bi$_{2}$Te$_{3}$-Cr$_{2}$Ge$_{2}$Te$_{6}$ heterostructure and the Ferromagnetic Insulator Cr$_{2}$Ge$_{2}$Te$_{6}$} 






\date{\today}


\author{Yao Tian}
\affiliation{Department of Physics \& Institute of Optical Sciences, University of Toronto, ON M5S 1A7, Canada}

\author{Huiwen Ji}
\affiliation{Department of Chemistry, Princeton University, Princeton, NJ 08540, USA}

\author{R. J. Cava}
\affiliation{Department of Chemistry, Princeton University, Princeton, NJ 08540, USA}

\author{L. D. Alegria}
\affiliation{Department of Physics, Princeton University, Princeton, New Jersey 08544, USA}

\author{J. R. Petta}
\affiliation{Department of Physics, Princeton University, Princeton, New Jersey 08544, USA}

\author{Kenneth S. Burch}
\email{ks.burch@bc.edu}
\affiliation{Department of Physics, Boston College 140 Commonwealth Ave Chestnut Hill, MA 02467-3804, USA}


\date{\today}

\begin{abstract}
Cr$_{2}$Ge$_{2}$Te$_{6}$ has been of interest for decades, as it is one of only a few ferromagnetic insulators. Recently, this material has been revisited due to its potential as a substrate for Bi$_{2}$Te$_{3}$, a topological insulator. This enables the possibility of studying the anomalous quantum Hall effect in topological insulators, and a route to novel spintronic devices. To probe the compatibility of these two materials, we use polarized variable temperature Raman microscopy to study Bi$_{2}$Te$_{3}$-Cr$_{2}$Ge$_{2}$Te$_{6}$  heterostructure as well as the phonon dynamics of Cr$_{2}$Ge$_{2}$Te$_{6}$.  We found the temperature dependence of the Cr$_{2}$Ge$_{2}$Te$_{6}$ phonons results primarily from anharmonicity, though a small magneto-elastic coupling is also observed. Our results confirm the potential of Cr$_{2}$Ge$_{2}$Te$_{6}$ as a substrate for topological insulators.

\end{abstract}

\maketitle

Cr$_{2}$Ge$_{2}$Te$_{6}$ is a particularly interesting material since it is in the very rare class of ferromagnetic insulators, and possesses a layered, nearly two dimensional structure.\cite{CGT_original} Recently this material has been revisited due to its small lattice mismatch with the topological insulator Bi$_{2}$Te$_{3}$, making it an ideal candidate as a substrate for novel devices.\cite{Huiwen_doc} Indeed, if Cr$_{2}$Ge$_{2}$Te$_{6}$  has a strong magnetic exchange with the Bi$_{2}$Te$_{3}$ film, it is predicted to reveal the anomalous quantum hall state.\cite{Hasan2010,Fu2009,BT_CGT_quantum_hall} This  may enable future topological and spintronic devices. Nonetheless this requires cooling below the T$_{c}$ of the Cr$_{2}$Ge$_{2}$Te$_{6}$, and thus it is important to establish the compatibility between the two materials at low temperatures. Furthermore, since the materials are van-der-waals bonded, magneto-elastic coupling between them is a likely exchange pathway. Thus understanding the phonon dynamics in Cr$_{2}$Ge$_{2}$Te$_{6}$ is crucial for determining its suitability as a potential substrate for topological devices.

Temperature dependent Raman scattering has been widely used to probe phonon dynamics as well as check the compatibility of compounds.\cite{compatible_Heterostructure_raman,Raman_Characterization_Graphene,Raman_graphene} Indeed, one can monitor the phonons and their respective symmetry to ensure no major change in the lattice occurs upon cooling, by controlling the polarization of the excitation source and detected photons.\cite{polarized_raman_study_of_BFO,polarized_raman_study_of_BFO_Beekman} Also,  the temperature dependence of the Raman scattering can uncover the magneto-elastic coupling.\cite{Pandey2013}
In this letter, we first check the compatibility between Bi$_{2}$Te$_{3}$ and Cr$_{2}$Ge$_{2}$Te$_{6}$ by performing Raman measurements at 293 K and 10 K on a Bi$_{2}$Te$_{3}$-Cr$_{2}$Ge$_{2}$Te$_{6}$ heterostructure. Then, we focus on the polarized temperature dependent Raman spectra of  Cr$_{2}$Ge$_{2}$Te$_{6}$, which reveal no major changes in the lattice at low temperatures. Furthermore we find the phonon temperature dependence is primarily governed by anharmonicity, though there are also small magneto-elastic effects. Thus lattice compatibility between Bi$_{2}$Te$_{3}$ and Cr$_{2}$Ge$_{2}$Te$_{6}$ is maintained at low temperatures, and the magneto-elastic effect may provide a route to coupling between the two.

The heterostructure in this study was grown by MOCVD deposition of Bi$_{2}$Te$_{3}$ on Cr$_{2}$Ge$_{2}$Te$_{3}$. The single crystal  Cr$_{2}$Ge$_{2}$Te$_{6}$ was grown with high purity elements heated to 700$^{o}$C. Detailed growth procedures of both materials can be found elsewhere.\cite{Huiwen_doc}  The Raman spectra on the heterostructure were taken in a backscattering configuration with a home-built Raman microscope.  A Tornado Hyperflux U1 spectrometer  with a cooled Andor iDus charge-coupled device(CCD) was used to record the data. Two Ondax Ultra-narrow-band Notch Filters were used to reject Rayleigh scattering. This also allows us to observe both Stokes and anti-Stokes  Raman shifts.  A solid-state 532 nm laser was used for the excitation. The sample was glued by silver paint onto a copper mount sitting in a commercial optical microscopy cryostat from Cryo Industries of America, Inc. A glass compensated Mitutoyo long working distance 50x (N.A.=0.42) objective was used for excitation as well as collection, resulting in a laser spot of 8 $\mu$m in diameter. The laser power was kept fairly low (300 $\mu$W) to avoid laser-induced heating. This was checked at 8 K by monitoring the anti-Stokes signal as the laser power was reduced. Once the anti-Stokes signal disappeared, the power was cut an additional $\approx 50\%$. The spectra were recorded with a polarizer in front of the spectrometer.

\begin{figure}
  \centering
  \includegraphics[width=8cm]{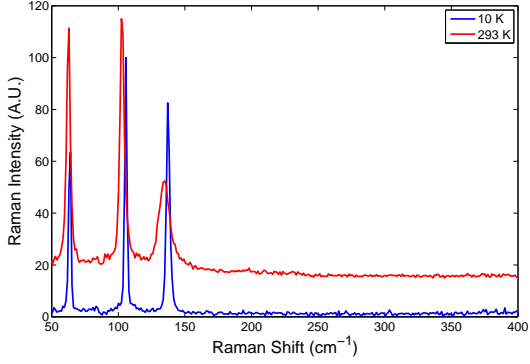}\\
  \caption{Raman spectra of Bi$_{2}$Te$_{3}$-Cr$_{2}$Ge$_{2}$Te$_{6}$ heterostructures in XX geometry at 293 K and 10 K. The two spectra  offset intentionally. The thickness of Bi$_{2}$Te$_{3}$ is about 100nm.}\label{BT_on_CGT}
\end{figure}

	The Raman spectra from the Bi$_{2}$Te$_{3}$-Cr$_{2}$Ge$_{2}$Te$_{6}$ heterostructure are shown in FIG.\ref{BT_on_CGT}. Only  the phonons of Bi$_{2}$Te$_{3}$ are observed, since the penetration depth of Bi$_{2}$Te$_{3}$ (20nm) is much smaller than the thickness of the film (100nm). The phonon peaks of Bi$_{2}$Te$_{3}$ at 293 K agree very well with the literature.\cite{Bi2Te3_Raman} As temperature decreases, the three phonons slightly harden and become sharper at 10 K, which we attribute to anharmonic renormalization. No new modes or dramatic changes in the spectra were observed. This suggests Cr$_{2}$Ge$_{2}$Te$_{6}$ and Bi$_{2}$Te$_{3}$ are well matched.

\begin{figure}
  \centering
  \includegraphics[width=8cm]{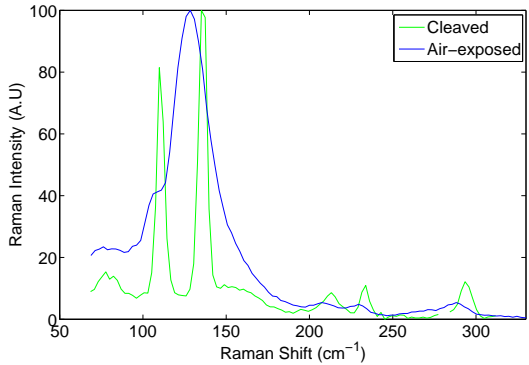}\\
  \caption{Raman spectra of Cr$_{2}$Ge$_{2}$Te$_{6}$ sample were taken in different conditions. All the spectra are taken at 300 K. Legend: the text states the condition of the sample.}\label{633_532_oldsample_raman}
\end{figure}

\begin{figure}
  \centering
  \includegraphics[width=8cm]{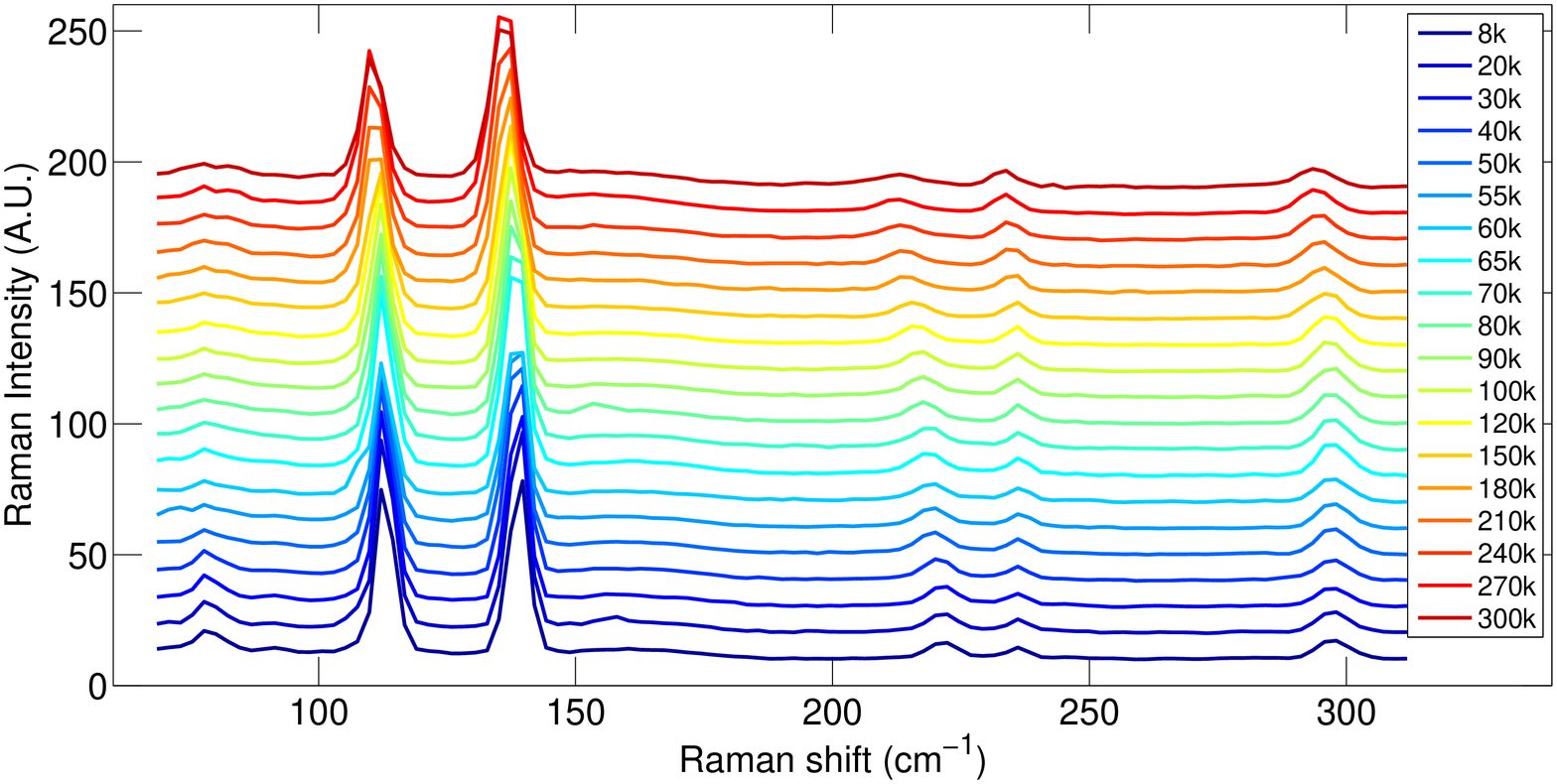}\\
  \caption{Temperature dependent collinear(XX) Raman spectra of Cr$_{2}$Ge$_{2}$Te$_{6}$ measured in the temperature range of 300 K $ - $ 8 K(Spectra are normalized to the mode at 138.8cm$^{-1}$).}\label{XX_temp}
\end{figure}

To confirm the compatibility of these two materials, we focused on the temperature dependent Raman spectra of Cr$_{2}$Ge$_{2}$Te$_{6}$. This analysis was complicated by the oxidation of the Cr$_{2}$Ge$_{2}$Te$_{6}$ surface. Indeed, it is well known that exposure to air can result in oxides on the surface that dominate the Raman spectra.\cite{Raman_aging_effect} The strong aging effect was also observed in Cr$_{2}$Ge$_{2}$Te$_{6}$. In FIG.\ref{633_532_oldsample_raman} we show the Raman spectra of air-exposed and freshly cleaved Cr$_{2}$Ge$_{2}$Te$_{6}$ samples. The air-exposed sample reveals fewer phonon modes, that are also quite broad, suggesting the formation of an oxide. 
A similar phenomena was also observed in other materials.\cite{Raman_amorphous_crystalline_transition_CGTfamily}
In FIG.\ref{XX_temp}, we show the colinearly polarized(XX) Raman specta of Cr$_{2}$Ge$_{2}$Te$_{6}$ from room temperature to 8 K. The sharpness of the phonon peaks indicates the high quality of the crystal. From the plot, we can see that there are six phonon peaks visible in the whole temperature range. At 8 K, their center frequencies are 78.7 cm$^{-1}$ ,113.0 cm$^{-1}$, 138.8 cm$^{-1}$ ,221.4 cm$^{-1}$, 236.1 cm$^{-1}$ and 297.4 cm$^{-1}$. While the small temperature induced shifts in the phonons suggests no major change in lattice constant, as discussed below, we can also search for symmetry breaking by looking at the polarization dependence.

From group theory analysis, the space group of single crystal Cr$_{2}$Ge$_{2}$Te$_{6}$  is R-3 (No.148). The point group is C$_{3i}$. There are six inequivalent irreducible representations in  the C$_{3i}$ point group. They are A$_{g}$, E$_{1g}$, E$_{2g}$, A$_{u}$, E$_{1u}$ and E$_{2u}$ respectively. All irreducible representations for the C$_{3i}$ point group are one dimensional. However E$_{1g}$ and E$_{2g}$ are inequivalent conjugate representations.  In this case, if the time reversal symmetry is not broken, the eigenstates of E$_{1g}$ and E$_{2g}$ representation are degenerate. This holds for  E$_{1u}$ and E$_{2u}$ representations as well. In the Cr$_{2}$Ge$_{2}$Te$_{6}$ unit cell, there are 10 atoms which gives 30-phonon branches at the $\Gamma$ point of the Brillouin  zone. The 30 phonon branches are $\Gamma_{acoustic}$ = A$_{u}$ + E$_{1u}$+ E$_{2u}$
and $\Gamma_{optical}$ = 5A$_{g}$ + 4A$_{u}$ + 5E$_{1g}$ + 4E$_{1u}$+ 5E$_{2g}$ + 4E$_{2u}$ respectively.   Because, the space group R-3 has inversion symmetry, theoretically  all the optical modes are either IR-active or Raman-active. The IR-active modes are  4A$_{u}$ + 4E$_{1u}$ + 4E$_{2u}$ and  the Raman-active modes are 5A$_{g}$ + 5E$_{1g}$ + 5E$_{2g}$. Here, the letter A means the phonon is non-degenerate and E means the phonon is double-degenerate for the reason explained above.
We expect to see 10 Raman-active modes, because the E$_{1g}$ and E$_{2g}$ mode are not distinguishable by energy. In the Raman spectra (Fig. \ref{XX_temp}), 6 modes were observed, the other four phonons might be too weak or out of our spectra range.\cite{Raman_ZnS3_modes_not_visible}

\begin{figure}
  \centering
  \includegraphics[width=8cm]{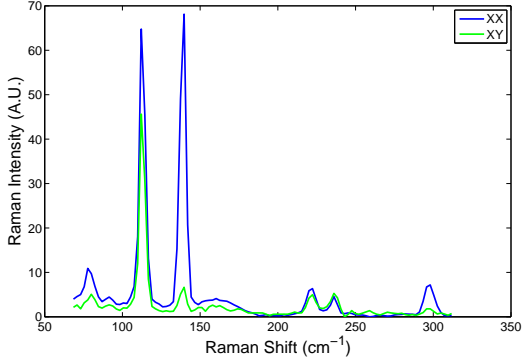}\\
  \caption{Normalized Raman Spectra of Cr$_{2}$Ge$_{2}$Te$_{6}$ in XX and XY geometry at 8 K.}\label{XY_temp}
\end{figure}

To analyze the polarization dependence, it is instructive to consider the relationship between the Raman tensor for a given mode (R) and the measured intensity I$_R$,
\begin{equation}\label{e}
  I_{R}=|\overrightarrow{\mathbf{e}}_{o}^{T}\mathbf{R}\overrightarrow{\mathbf{e}}_{i}|^{2}
\end{equation}
where $\overrightarrow{\mathbf{e_{o}}}$,$\overrightarrow{\mathbf{e_{i}}}$ are the polarization of the out-going and the incoming photon respectively. The Raman tensors of phonon of A$_{g}$, E$_{1g}$ and E$_{2g}$ symmetry are shown below.
\[A_{g}=\left(
  \begin{array}{ccc}
    a & 0 & 0 \\
    0 & a & 0 \\
    0 & 0 & b \\
  \end{array}
\right)
E_{1g}=\left(
  \begin{array}{ccc}
    c & d & e \\
    d & -c & f \\
    e & f & 0 \\
  \end{array}
\right)\]
\[
E_{2g}=\left(
         \begin{array}{ccc}
           d & -c & -f \\
           -c & -d & e \\
           -f & e & 0 \\
         \end{array}
       \right)
\]
From the Raman tensor, we know that all three modes should be visible in the colinear polarized(XX) geometry and the  A$_{g}$ modes should vanish in crossed polarized(XY) geometry (note the absence of off diagonal elements in A$_{1g}$). To gain insight into the symmetry of the modes, we compare the spectra taken at 8 K in XX and XY configurations. As can be seen from FIG.\ref{XY_temp}, only the two modes located at 138.8 cm$^{-1}$ and  297.4 cm$^{-1}$ vanish in XY configuration. Therefore these two modes are of A$_{g}$ symmetry, and the other four modes are of E$_{g}$ symmetry. Furthermore we confirmed the symmetry is maintained at all temperature measured (not shown).

\begin{figure}
  \centering
  \includegraphics[width=10cm]{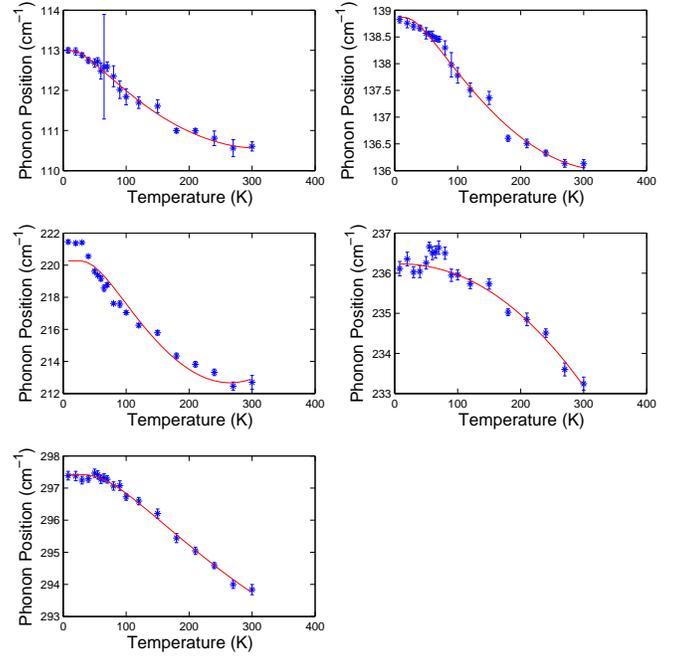}\\
  \caption{Temperature dependence of the shifts of phonons. For all but the mode centered near 222 cm$^{-1}$ the temperature dependence is well explained by anharmonic effects. However, below T$_{c}$ this mode also reveals additional shifting due to magneto-elastic coupling.}\label{Phonon_Temperature_position}
\end{figure}

In order to analyse the phonon temperature dependent behavior, we used the Lorentz model to fit each individual phonon peak at all temperatures. The mode located at 78.7 cm$^{-1}$ has relatively low intensity between 100 K to 300 K, and thus was excluded from our fitting procedure. We show the resulting phonon frequencies in FIG.\ref{Phonon_Temperature_position}, which all harden as temperature is decreased.  This result is not surprising since the anharmonic interaction leads to a renormalization of the phonon self energy by both decay and fusion processes. However, the fusion process, where two or more phonons fuse into a zone-center phonon, is usually very slow due to the low population of phonons at low temperature.\cite{Anharmonic_phonon_scattering} This indicates that the scattering is almost exclusively governed by the decay process. Ultimately, this results in a temperature dependent renormalization of the real (frequency) and imaginary (inverse lifetime) parts of the phonon self-energy. To further quantitatively analyze the anharmonic effect, we use a simple model which takes the cubic and quartic anharmonic decay into account.  In this model, the temperature dependence of the phonon frequency is described by the formula,\cite{PhysRevB.28.1928}
\begin{eqnarray}
   &\omega(T)=\omega_{0}+\Delta\omega(T)  \\
   &\Delta\omega(T)=C(1+\frac{2}{e^{x}-1})+D(1+\frac{3}{e^{y}-1}+\frac{3}{(e^{y}-1)^{2}})\label{Phonon_Temperature_equation}
\end{eqnarray}
where $\omega_{0}$  is the phonon position at 0 K, T is the temperature, C,D are constants determined by the cubic and quartic anharmonicity respectively, x = $\hbar\omega_{0}$/2$k_{B}T$ and y = $\hbar\omega_{0}$/3$k_{B}T$. The first term in equation \ref{Phonon_Temperature_equation} describes the optical phonon decay into two phonons with opposite momenta and half the energy of the original phonon. The second term describes  the optical phonon decay into three phonons with a third of the energy of the optical phonon. The results of fitting, the temperature dependence of the phonon shifts with equation \ref{Phonon_Temperature_equation} are shown in FIG.\ref{Phonon_Temperature_position}. From the plot we can see the anharmonic interaction model works reasonably well for all but the 212 cm$^{-1}$ mode. The resulting parameters for the anharmonic interaction are shown in Table.\ref{fit_result}. From the table we can see that the four-phonon interaction is much weaker, as is expected.\cite{PhysRevB.28.1928} Below T$_{c}$,  the mode at 212 cm$^{-1}$ shifts more rapidly than the than the prediction of the anharmonic model. This is likely caused by magneto-elastic coupling in Cr$_{2}$Ge$_{2}$Te$_{6}$.\cite{Pandey2013,F_phonon_coupling1}

In summary, we verified the compatibility of Bi$_{2}$Te$_{3}$-Cr$_{2}$Ge$_{2}$Te$_{6}$ heterostructure by Raman spectroscopy at 10 K and 293 K. The stability of the Cr$_{2}$Ge$_{2}$Te$_{6}$ single crystal was also studied using polarized variable temperature Raman spectroscopy. Six phonons were observed and they were explained by anharmonic interaction. A weak sign of the magneto-elastic coupling has also been observed, suggesting an additional pathway for magnetic coupling between the two materials. Thus we have shown the utility of using Cr$_{2}$Ge$_{2}$Te$_{6}$ as a substrate for future devices.

Work at the University of Toronto was supported by NSERC, CFI, and ORF.

\begin{table}[ht]
  \caption{Anharmonic interaction parameters}
  \centering
  \begin{tabular}{c c c c c c}
  \hline\hline
  C & Error & D & Error & $\omega_{0}$ & Error\\[0.5ex]
  \hline
    -0.835 & 0.138 & 0.032 & 0.0102 & 113.0 & 0.200 \\
    -1.19 & 0.19 & 0.0525 & 0.0168 & 138.9 & 0.250 \\
    -9.70 & 3.06 & 0.865 & 0.403 & 220.2 & 2.95 \\
    -1.3 & 0.171 & 0.0203 & 0.0097 & 236.5 & 0.300 \\
    -2.93 & 0.763 & 0.147 & 0.120 & 297.2 & 0.700 \\
  \hline\label{fit_result}
\end{tabular}
\label{Anharmonic_fit_data}
\end{table}


%




%

\
\end{document}